\documentclass[aps,10pt,showpacs,showkeys,superscriptaddress,footinbib,twocolumn]{revtex4}
\usepackage{graphicx,epstopdf}
\usepackage{color} 

\newcommand{\im}{\mathrm{Im}}

\begin{document}
\itemindent-3mm
\title{Comment on: Non-perturbative finite T broadening of the rho
  meson\\
 and dilepton emission in heavy ion-collisions\\ 
 {\small by J. Ruppert   and T. Renk\\ 
 Phys.\ Rev.\ C {\bf 71} (2005) 064903; nucl-th/0412047} } 

\author{Felix Riek}\thanks{e-mail: F.Riek@gsi.de}
\affiliation{Gesellschaft
 f\"ur Schwerionenforschung mbH, Planckstr. 1,
D-64291 Darmstadt, Germany}
\author{Hendrik van Hees}\thanks{e-mail: hees@comp.tamu.edu}
\affiliation{Cyclotron Institute, Texas A\&M University, 
College Station, Texas 77843-3366, USA}
\author{J\"orn Knoll}\thanks{e-mail: J.Knoll@gsi.de}
\affiliation{Gesellschaft
 f\"ur Schwerionenforschung mbH, Planckstr. 1, 
D-64291 Darmstadt, Germany}

\begin{abstract}
  In this comment we point out several problems concerning kinematical
  singularities which are encountered in the calculation of the dilepton
  rates in~\cite{Ruppert:2004yg}. We also comment on
  the method introduced in \cite{vanHees:2000bp} and further used in
  refs.
  ~\cite{Riek:2004kx,Renk:2006dt,Renk:2006ax,Ruppert:2004yg}.
\end{abstract}

\date{\today}
\maketitle

  In this publication~\cite{Ruppert:2004yg} the authors
  find a surprising broadening of the $\rho$-meson solely due to its
  interactions with pions.  Since this is in contrast with expectations
  from low-energy QCD and the implied Goldstone-boson nature of the
  pion~\cite{Chiral} we tried to find the reason for this behavior.

Our reanalysis leads us to the conclusion that the projection
method introduced in~\cite{Ruppert:2004yg} in order to restore the four
transversality of the $\rho$-meson spectral function,
  $A^{\mu\nu}$, suffers from serious problems with kinematical
singularities. As shown below these singularities lead to a
\emph{spurious and unphysical} massless mode of the $\rho$-meson, in
turn leading to a further broadening of the pion modes and through the
self-consistence finally to a strong broadening of the $\rho$-meson.
  In recent studies ~\cite{Renk:2006dt,Renk:2006ax} the
  authors alternatively used the projection method introduced by two of
  us \cite{vanHees:2000bp} and found a good agreement with the dimuon
  data of NA60~\cite{NA60} on the basis of their collision-dynamic
  model~\cite{Ruppert:2004yg}.  We point out that also in this
  projection method there are ambiguities in the calculation.

\textbf{The analysis in detail:} The authors use a $\Phi$-derivable
self-consistent Dyson-resummation scheme to evaluate self energies
of vector mesons, which, a priori, is a promising method. However,
  it suffers from the violation of Ward-Takahashi identities at the
two-point and higher-order vertex functions level, leading to a
violation of current conservation within the self-consistent propagators
although the expectation value of the current is conserved.  This leads
to the artificial excitation of the unphysical four-dimensionally
longitudinal mode of the vector meson and thus to a violation of
unitarity.

In order to cure this defect the authors employed a naive projection
scheme. In any iteration step of the self-consistent scheme it simply
cuts off the undesired four-longitudinal components of the polarization
tensor, $\Pi^{\mu\nu}(q)$, of the $\rho$-meson.
\begin{figure}[t]
\includegraphics[width=0.8\linewidth]{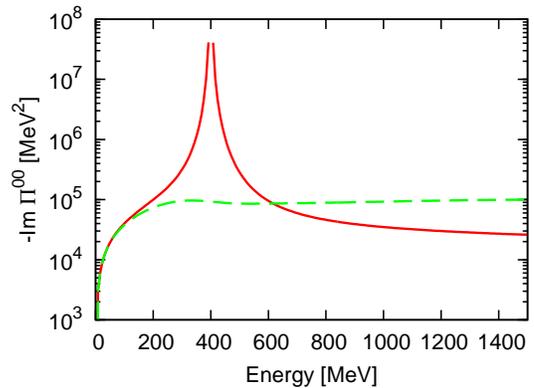}
\caption{Time-time component of the $\rho$-meson polarization tensor at
  $q=400$~MeV and $T=160$~MeV before (dashed) and after (full line) the
  projection method of Ruppert-Renk.\\[-0.8cm]}
\label{RR-Sigma-00}
\end{figure}
Since the projectors, however, are singular on the light-cone, the
spatially longitudinal component of the polarization tensor,
\begin{equation}
  \mathrm{Im}\;\Pi_L^{\mu\nu}(q) \stackrel{q^2\rightarrow 0}{\longrightarrow}
  \epsilon(q)\frac{q^{\mu}
    q^{\nu}}{(q^2)^2},
\end{equation}
becomes divergent on the light cone, cf. Fig. \ref{RR-Sigma-00}. 
Here $\epsilon(q)$ is a measure of
the violation of four transversality on the light cone, since proper
four transversality requires $\lim_{q^2\to 0} \epsilon(q)=0$!  The
occurrence of this singularity was already stated by the authors
themselves~\cite{Ruppert:2004yg}, though qualified as harmless!
However, it strongly violates analyticity requirements, since
\begin{figure}[t]
\vspace*{-1cm}
\hspace*{-.2cm}\includegraphics [width=0.95\linewidth]{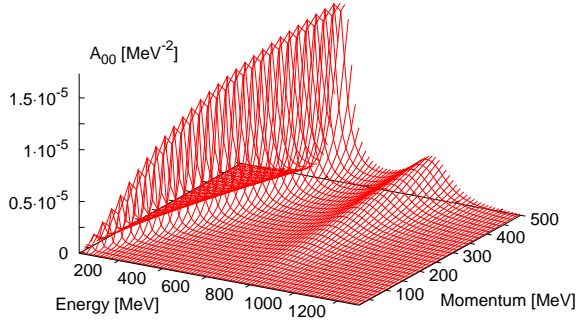}
\vspace*{-0.6cm}
\caption{Time-time component $A^{00}(q)$ of the $\rho$-meson spectral
  function at $T=160$~MeV as a function of energy and three momentum for
  the Ruppert-Renk projection method.\\[-0.8cm]}
\label{RR-A-00-3d}
\end{figure}
\begin{itemize}
\item[a)] $\Pi^{\mu\nu}(q)$ is given by the space-time Fourier
  transformation of the corresponding current-current correlator
  $\langle J^{\mu}(x) J^{\nu}(0)\rangle$. Thus, apart from UV
  regularizations the four-momentum Fourier transformation of
  $\Pi^{\mu\nu}(q)$ must exist, not to mention serious value constraints
  on $\Pi^{\mu\nu}$ arising from sum rules such as the
  Thomas-Reiche-Kuhn sum rule, the f-sum rule, or Weinberg's sum rule
  which all would diverge by this construction!
\item[b)] the corresponding physically relevant Lorentz components of
  the $\rho$-meson spectral-function can simply be estimated
  analytically,
\begin{equation}
  A_L^{\mu\nu}(q)\stackrel{q^2 \rightarrow 0}{\longrightarrow} \frac{2
    \epsilon q^\mu q^\nu }{4m_\rho^4 \vec{q}^2(q^0-|\vec{q}|)^2+\epsilon^2},
\end{equation}
in the vicinity of the light-cone. There these components show a strong
peak, c.f. the result for $A^{00}$ in Figs.~\ref{RR-A-00-3d}
obtained from our numerical repetition of the
Ruppert-Renk method.
\end{itemize}
Note that $A^{00}(q)=\frac{{\vec q}^2}{q^2} A_L(q)$ drops to zero at
vanishing spatial momentum $\vec q$.  This light-cone structure
represents a zero-mass mode with amazing stamina. This fictitious mode
always emerges unless the unprojected tensor is \emph{exactly}
four-transversal on the light-cone. It has the remarkable feature that
for any given momentum $\vec{q}$, its energy-weighted integral strength
(obtained from the residue) is about identical to the resonance strength
integrated across $m_\rho$! Similar conclusions hold for the spatial
components of $A_L$.
 
Two of us (HvH and JK)~\cite{vanHees:2000bp} suggested an alternative
method for the construction of a four-transversal polarization tensor
which definitely avoids the above stated light-cone singularity, since
there $\Pi_L(q)$ vanishes by construction. For details we refer to
refs.~\cite{vanHees:2000bp,Riek:2004kx}. Here one discards the
self-consistently obtained time components $\Pi^{00}(q)$ and
$\Pi^{0i}$, $\Pi^{i0}$, since due to the conservation law they involve
an infinite relaxation time which is known to escape a reliable
treatment in self-consistent schemes at finite loop order.
Therefore the full tensor is constructed solely from the
self-consistently obtained spatial components $\Pi^{ik}$ such that
$\Pi^{\mu \nu}$ becomes exactly four-dimensionally transversal. It
should be mentioned though that due to the $1/q_0^2$ factor in the
\begin{figure}[t]
\vspace*{-1cm}
\hspace*{-.2cm}\includegraphics [width=0.95\linewidth]{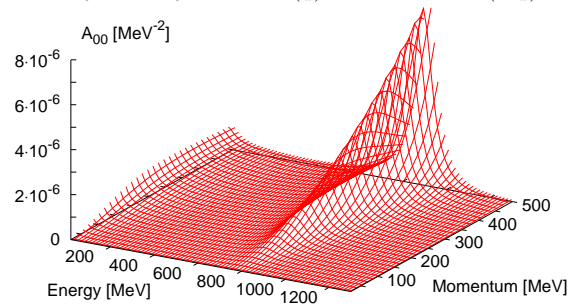}
\vspace*{-0.6cm}
\caption{Same as in Fig.~\ref{RR-A-00-3d} for the method by van Hees
  and Knoll. \\[-1.0cm] }
\label{vHK-A-00-3d}
\end{figure}
construction of $\Pi^{00}(q)=q_iq_k\Pi^{ik}/q_0^2$, this method may lead
to a less controlled determination of $A^{00}$ close to vanishing energy
$q_0=0$. Even though we expect contributions arising
  from classical random scattering~\cite{Knoll:1995nz} (the hight $T$
  limit of Landau damping), which indeed strongly peak close to $q_0=0$
  we point out that they are essentially uncontrolled.  Comparing the
  numerical result given in Fig.~\ref{vHK-A-00-3d}, this component may
  look tiny (possibly due to the antisymmetry $\im \;
    A^{00}(q)=-\im \; A^{00}(-q)$, which suppresses the components near
    $q_0=0$) as compared to the artificial light-cone mode of the
  Ruppert-Renk method, Fig.~\ref{RR-A-00-3d} (note the differences in
  the ordinate scales) but additional clarification is mandatory.\\[-8mm]

\section*{Conclusions}
The zero-mass mode of the $\rho$-meson produced by the Ruppert-Renk
projection method rests on a kinematical singularity and is therefore
completely unphysical. In the self-consistent scheme it provides a
strong new decay mode for the $\pi\pi\rho$-coupling, which in turn
significantly broadens the pion spectral function, and finally leads to
the stated broadening of the $\rho$-meson! Since the projection strategy
of van Hees and Knoll was also used in the studies
\cite{Renk:2006ax,Renk:2006dt,Riek:2004kx} we will carefully
reinvestigate the scheme and hopefully achieve a concept that also
complies with the sum-rule constraints for the polarization tensor,
before we subject the method to a quantitative comparison with data.
We strongly support the statement made by the authors in
  \cite{Renk:2006ax} that the current status of the model does not allow
  to conclude that nonperturbative $\rho-\pi$ interactions are the main
  mechanism for the broadening observed in the NA60 data.

We acknowledge clarifying discussions with S. Damjanovic, H. Specht, B.
Friman, R. Rapp, J. Wambach, D. Rischke, J. Ruppert and
  T. Renk

\end{document}